\RequirePackage{lineno}
\documentclass[superscriptaddress,showpacs,preprintnumbers,amsmath,amssymb]{revtex4}

\def\GeVcSq {{\rm (GeV/}c)^2}

\usepackage{graphicx}
\usepackage{dcolumn}
\usepackage{booktabs}
\usepackage{bm}
\usepackage{color}
\usepackage{ulem}

\begin{document}
\preprint{}
\title{Single Spin Asymmetry $A_N$ 
in Polarized Proton-Proton Elastic Scattering \\ at $\sqrt{s}=200$~GeV} 
\affiliation{AGH University of Science and Technology, Cracow, Poland}
\affiliation{Argonne National Laboratory, Argonne, Illinois 60439, USA}
\affiliation{University of Birmingham, Birmingham, United Kingdom}
\affiliation{Brookhaven National Laboratory, Upton, New York 11973, USA}
\affiliation{University of California, Berkeley, California 94720, USA}
\affiliation{University of California, Davis, California 95616, USA}
\affiliation{University of California, Los Angeles, California 90095, USA}
\affiliation{Universidade Estadual de Campinas, Sao Paulo, Brazil}
\affiliation{Central China Normal University (HZNU), Wuhan 430079, China}
\affiliation{University of Illinois at Chicago, Chicago, Illinois 60607, USA}
\affiliation{Cracow University of Technology, Cracow, Poland}
\affiliation{Creighton University, Omaha, Nebraska 68178, USA}
\affiliation{Czech Technical University in Prague, FNSPE, Prague, 115 19, Czech Republic}
\affiliation{Nuclear Physics Institute AS CR, 250 68 \v{R}e\v{z}/Prague, Czech Republic}
\affiliation{University of Frankfurt, Frankfurt, Germany}
\affiliation{Institute of Physics, Bhubaneswar 751005, India}
\affiliation{Indian Institute of Technology, Mumbai, India}
\affiliation{Indiana University, Bloomington, Indiana 47408, USA}
\affiliation{Alikhanov Institute for Theoretical and Experimental Physics, Moscow, Russia}
\affiliation{University of Jammu, Jammu 180001, India}
\affiliation{Joint Institute for Nuclear Research, Dubna, 141 980, Russia}
\affiliation{Kent State University, Kent, Ohio 44242, USA}
\affiliation{University of Kentucky, Lexington, Kentucky, 40506-0055, USA}
\affiliation{Institute of Modern Physics, Lanzhou, China}
\affiliation{Lawrence Berkeley National Laboratory, Berkeley, California 94720, USA}
\affiliation{Massachusetts Institute of Technology, Cambridge, MA 02139-4307, USA}
\affiliation{Max-Planck-Institut f\"ur Physik, Munich, Germany}
\affiliation{Michigan State University, East Lansing, Michigan 48824, USA}
\affiliation{Moscow Engineering Physics Institute, Moscow Russia}
\affiliation{National Institute of Science and Education and Research, Bhubaneswar 751005, India}
\affiliation{Ohio State University, Columbus, Ohio 43210, USA}
\affiliation{Old Dominion University, Norfolk, VA, 23529, USA}
\affiliation{Institute of Nuclear Physics PAN, Cracow, Poland}
\affiliation{Panjab University, Chandigarh 160014, India}
\affiliation{Pennsylvania State University, University Park, Pennsylvania 16802, USA}
\affiliation{Institute of High Energy Physics, Protvino, Russia}
\affiliation{Purdue University, West Lafayette, Indiana 47907, USA}
\affiliation{Pusan National University, Pusan, Republic of Korea}
\affiliation{University of Rajasthan, Jaipur 302004, India}
\affiliation{Rice University, Houston, Texas 77251, USA}
\affiliation{Universidade de Sao Paulo, Sao Paulo, Brazil}
\affiliation{University of Science \& Technology of China, Hefei 230026, China}
\affiliation{Shandong University, Jinan, Shandong 250100, China}
\affiliation{Shanghai Institute of Applied Physics, Shanghai 201800, China}
\affiliation{SUBATECH, Nantes, France}
\affiliation{Temple University, Philadelphia, Pennsylvania, 19122}
\affiliation{Texas A\&M University, College Station, Texas 77843, USA}
\affiliation{University of Texas, Austin, Texas 78712, USA}
\affiliation{University of Houston, Houston, TX, 77204, USA}
\affiliation{Tsinghua University, Beijing 100084, China}
\affiliation{United States Naval Academy, Annapolis, MD 21402, USA}
\affiliation{Valparaiso University, Valparaiso, Indiana 46383, USA}
\affiliation{Variable Energy Cyclotron Centre, Kolkata 700064, India}
\affiliation{Warsaw University of Technology, Warsaw, Poland}
\affiliation{University of Washington, Seattle, Washington 98195, USA}
\affiliation{Wayne State University, Detroit, Michigan 48201, USA}
\affiliation{Yale University, New Haven, Connecticut 06520, USA}
\affiliation{University of Zagreb, Zagreb, HR-10002, Croatia}

\author{L.~Adamczyk}\affiliation{AGH University of Science and Technology, Cracow, Poland}
\author{G.~Agakishiev}\affiliation{Joint Institute for Nuclear Research, Dubna, 141 980, Russia}
\author{M.~M.~Aggarwal}\affiliation{Panjab University, Chandigarh 160014, India}
\author{Z.~Ahammed}\affiliation{Variable Energy Cyclotron Centre, Kolkata 700064, India}
\author{A.~V.~Alakhverdyants}\affiliation{Joint Institute for Nuclear Research, Dubna, 141 980, Russia}
\author{I.~Alekseev}\affiliation{Alikhanov Institute for Theoretical and Experimental Physics, Moscow, Russia}
\author{J.~Alford}\affiliation{Kent State University, Kent, Ohio 44242, USA}
\author{C.~D.~Anson}\affiliation{Ohio State University, Columbus, Ohio 43210, USA}
\author{D.~Arkhipkin}\affiliation{Brookhaven National Laboratory, Upton, New York 11973, USA}
\author{E.~Aschenauer}\affiliation{Brookhaven National Laboratory, Upton, New York 11973, USA}
% ...

\collaboration{STAR Collaboration}\noaffiliation



\homepage{}
\begin{abstract}
We report a high precision measurement of the  transverse single spin asymmetry $A_N$
at the center of mass energy $\sqrt{s}=200$ GeV in elastic proton-proton scattering
by the STAR experiment  at RHIC. 
The $A_N$ was measured in the four-momentum transfer squared $t$ range $0.003 \leqslant |t| \leqslant 0.035$ $\GeVcSq$,
the region of a significant interference between the electromagnetic
and hadronic scattering amplitudes. The measured values of $A_N$ and its $t$-dependence
are consistent with a vanishing hadronic spin-flip amplitude,
thus providing strong constraints on the ratio of the single spin-flip to the non-flip amplitudes. Since the hadronic amplitude is dominated by the Pomeron amplitude at this $\sqrt{s}$, we conclude that this measurement addresses the question about the 
presence of a hadronic spin flip due to the Pomeron exchange in polarized proton-proton elastic scattering. 
\end{abstract}
\pacs{13.85.Dz and 13.88.+e}
\keywords{Polarization, Elastic Scattering}
\maketitle
\section{\label{sec:intro}Introduction\protect\\}
High energy diffractive hadronic scattering at small values of four-momentum transfer squared $t$, is dominated by
an exchange of the Pomeron trajectory, a color-singlet object with the 
quantum numbers of the vacuum~\cite{barone,donnachie}.
The calculation of cross-sections for small-$t$ scattering requires a non-perturbative approach in QCD and its theoretical treatment is
still being developed.  The experimental data therefore provide significant constraints
for theoretical approaches and models~\cite{buttimore,trueman3}.
The coupling of the Pomeron to the nucleon spin is of special interest since it is predicted to be
sensitive to the internal dynamics of the nucleon~\cite{buttimore,trueman3}.
Studies of the spin dependence of proton-proton (pp) scattering at small momentum transfers and at
the highest energies presently available at RHIC offer an opportunity to reveal important
information on the nature of the Pomeron.

There are several theoretical approaches which predict non-zero spin-dependent Pomeron amplitudes  for elastic scattering.
Examples include an approach in which the Pomeron-proton coupling is modeled via two-pion
exchange~\cite{pumplin}, an 
impact picture model assuming that the spin-flip contribution is sensitive to the impact parameter distribution of matter in
a polarized proton~\cite{bourrely}, and
a model which treats the
Pomeron helicity coupling analogously to that of the isoscalar anomalous magnetic moment of
the nucleon~\cite{ryskin}.
Still another approach assumes a diquark enhanced picture of the proton~\cite{diquark}, in which
a non-zero spin-flip amplitude may arise if the proton wave function is dominated by an 
asymmetric
configuration, such as a quark-diquark.

Here we present a high precision measurement of the transverse single spin asymmetry $A_N$ in 
elastic scattering of polarized protons at 
$\sqrt{s} = 200 \:\rm{GeV}$ in the $t$-range $0.003 \leqslant |t| \leqslant 0.035$ $\GeVcSq$ by the STAR
experiment~\cite{starnim}  at RHIC.
The single spin asymmetry $A_N$ is defined as the left-right cross-section asymmetry with respect to the
transversely polarized proton beam. In this range of $t$,  $A_N$ originates predominantly from the 
interference between electromagnetic (Coulomb) spin-flip and hadronic (nuclear) non-flip 
amplitudes \cite{buttimore}. However, it was realized that $A_N$ in the Coulomb-nuclear 
interference (CNI) region is also a sensitive probe  of the hadronic
spin-flip amplitude \cite{diquark}, which will be discussed in more detail in Section~\ref{sec:hadronic}.

A previous measurement of $A_N$ in a similar $t$-range and the same $\sqrt{s}$, but with limited
statistics, has been reported by the pp2pp collaboration \cite{pp2ppAN}.
Other measurements of $A_N$ performed at small $t$ were obtained at significantly lower energies.
They include high precision results from the RHIC 
polarimeters obtained at $\sqrt{s} = 6.8 - 21.7~\rm{GeV}$  for elastic proton-proton \cite{jetcal,alekseev,bazilevsky} and
proton-carbon~\cite{carboncal} scattering, as well as earlier results from the BNL AGS for
pC scattering~\cite{e950} at $\sqrt{s} = 6.4~\rm{GeV}$ and from FNAL E704
for pp scattering~\cite{e704} at $\sqrt{s} = 19.4~\rm{GeV}$.

The combined analysis of all results, which covers a wide energy range and different targets, will help to disentangle 
contributions of various exchange mechanisms relevant for elastic scattering in the forward 
region \cite{trueman}. In particular, such an analysis will allow us to extract information on the spin dependence of the diffractive mechanism which dominates at high energies.

\section{\label{sec:hadronic}Hadronic spin-flip amplitude in elastic collisions}
Elastic scattering of two protons is described by five independent helicity
amplitudes: two helicity conserving ($\phi_1$ and $\phi_3$), two double helicity-flip ($\phi_2$ and $\phi_4$), 
and one single helicity-flip amplitude ($\phi_5$) -- see~\cite{buttimore} for definitions.
At very high $\sqrt{s}$, such as available at RHIC, and very small
 $|t| <$ 0.05~$\GeVcSq$, the proton mass $m$ can be neglected with respect to $\sqrt{s}$ and $t$ can be neglected 
with respect to $m$, which simplifies kinematical
factors in the following formulas. The elastic spin-averaged cross-section is
given by:
\begin{linenomath}
\begin{equation}
\frac{d\sigma}{dt} = \frac{2\pi}{s^2} {(|\phi_1|^2 + |\phi_2|^2 + |\phi_3|^2 + |\phi_4|^2 + 4 |\phi_5|^2)} \;,
\end{equation}
\end{linenomath}
while the single spin-flip
amplitude $\phi_5$ gives rise to the single spin asymmetry, $A_N$, through
interference with the remaining amplitudes:
\begin{linenomath}
\begin{equation}
A_N \frac{d\sigma}{dt}    = -\frac{4\pi}{s^2} {\rm Im}\:\{\phi_5^*(\phi_1+\phi_2+\phi_3-\phi_4)\} \;.
\end{equation}
\end{linenomath}
Each of the amplitudes consists of Coulomb and hadronic contributions: $\phi_i=\phi_i^{\rm em} + \phi_i^{\rm had}$,
with the electromagnetic one-photon exchange amplitudes  $\phi_i^{\rm em}$ described by QED using the measured anomalous magnetic moment of the proton.~\cite{buttimore78}.
The optical theorem relates the hadronic amplitudes to the total cross-section:
\begin{linenomath}
\begin{equation}
\sigma_{\rm total}              = \frac{4\pi}{s}{\rm Im\:}(\phi_1^{\rm had} + \phi_3^{\rm had})|_{t=0}\;,
\end{equation}
\end{linenomath}
which provides an important constraint on the parameterization of these dominant
helicity conserving hadronic amplitudes.

The contribution of the two double spin-flip hadronic amplitudes $\phi_2^{\rm had}$ and $\phi_4^{\rm had}$ to the asymmetry $A_N$ 
is small, as indicated by both experimental results~\cite{doublespin,spin2010} and theoretical predictions~\cite{trueman2}.
Thus, the main contribution to $A_N$ is given by:
\begin{linenomath}
\begin{equation}
A_N \frac{d\sigma}{dt} = -\frac{8\pi}{s^2} {\rm Im\:}(\phi_5^{{\rm em}*}\phi_+^{\rm had} + \phi_5^{{\rm had}*}\phi_+^{\rm em}) \;,
\end{equation}
\end{linenomath}
where $\phi_+=(\phi_1+\phi_3)/2$. 

The parametrization of $\phi_5^{\rm had}$ is usually done
in terms of 
\mbox{$\phi_+^{\rm had}\:$: $\phi_5^{\rm had}(s,t) = (\sqrt{-t}/m){\:\cdot\:}r_5(s){\:\cdot\:}{\rm Im\:}\phi_+^{\rm had}(s,t)$,}
where $m$ is the proton mass. Thus $r_5$ is the measure of the ratio of the hadronic single spin-flip amplitude ($\phi_5$) to
hadronic single non-flip amplitudes ($\phi_1$ and $\phi_3$). Using this parametrization the following representation of $A_N$
can be derived~\cite{buttimore}:

\begin{widetext}
\begin{linenomath}
\begin{equation}
A_N = \frac {\sqrt{-t}}{m} \: \frac {[\kappa (1 - \rho \:\delta) + 2 (\delta \:{\rm{Re}} \:r_5 - {\rm{Im}} \:r_5)] \frac{t_c}{t} - 2 ({\rm{Re}} \:r_5 - \rho \:{\rm{Im}} \:r_5)}{ (\frac{t_c}{t})^2 - 2 (\rho + \delta)\frac{t_c}{t} + (1 + \rho ^2)}\,,
\label{cnicurve}
\end{equation}
\end{linenomath}
\end{widetext}
where $t_c = -8 \pi \alpha /\sigma _{\rm total}$, $\kappa$ is the anomalous
magnetic moment of the proton, $\rho=\rm{Re\:}\phi_+/\rm{Im\:}\phi_+$ is the ratio of the real to imaginary
parts of the non-flip elastic amplitude, and $\delta $ is the relative phase between 
the Coulomb and hadronic amplitudes~\cite{buttimore}:
\begin{linenomath}
\begin{equation}
\delta = \alpha\: \ln\:\frac{2}{|t|(B+8/\Lambda ^2)} - \alpha\:\gamma \;,
\label{delta}
\end{equation}
\end{linenomath}
where $B$ is the slope of the forward peak in elastic scattering, 
$\alpha = 1/137$ is the fine structure constant, $\gamma = 0.5772$ is Euler's constant,
and $\Lambda ^2 = 0.71 \:\GeVcSq$.

\section{Detection of elastic proton-proton collisions at RHIC}
\begin{figure}
\includegraphics[width=130mm]{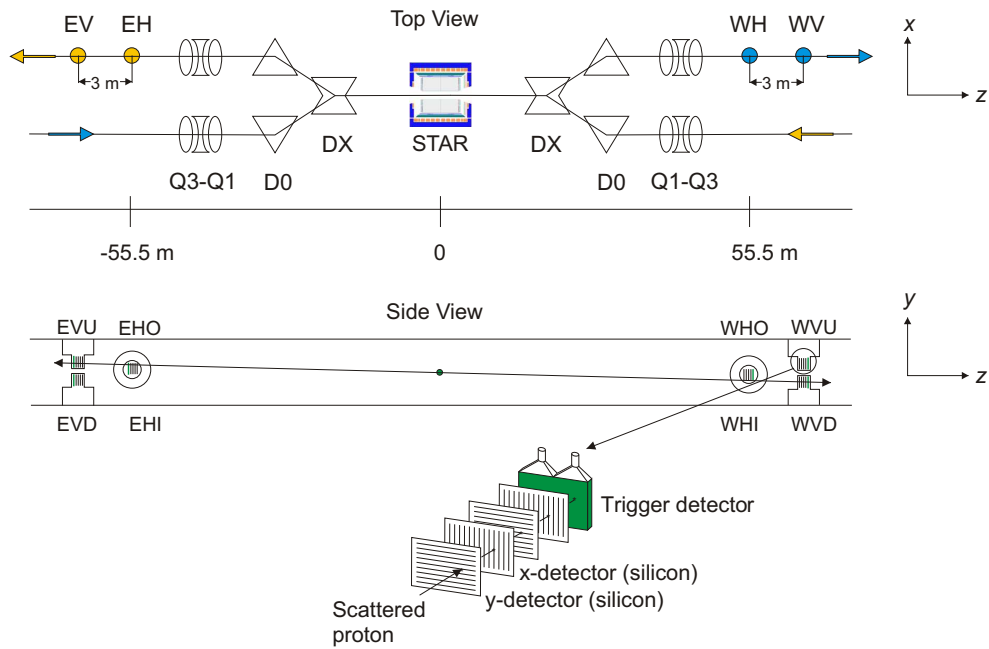}
\caption{(color online) The layout of the experiment. The Roman Pot stations are located on both sides of the STAR IP.
The positive $z$ direction is defined along the outgoing ``Blue" beam (the West direction).
Positive $y$ is pointing up and positive $x$ is pointing away from the center of the RHIC ring. The detectors are placed on the outgoing beams. The figure is not to scale.}
\label{fig:setup}
\end{figure}

The protons, which scatter elastically at small angles ($\lesssim$ 2 mrad), follow the optics of the RHIC 
magnets and are detected by a system of detectors placed
close to the beam inside movable vessels known as ``Roman Pots" (RPs)~\cite{pp2ppNIM}.
The Roman Pot stations are located on either side of the STAR interaction point (IP)
at 55.5 m and 58.5 m with horizontal and vertical insertions of the detectors, respectively.
The coordinate system of the experiment is described in Fig.~\ref{fig:setup}.
There are eight Roman Pots, four on each side of the IP. Four approach the beam horizontally WHI, WHO (EHI, EHO) and 
four approach the beam vertically WVU, WVD (EVU, EVD) as shown in Fig.~\ref{fig:setup}.
The location of the RPs  was optimized so that, combined with proper accelerator magnet settings,
it provides so-called ``parallel-to-point focusing", i.e. the $(x,y)$ position of the scattered protons at 
the RPs depends almost exclusively on their scattering angles 
and is nearly insensitive to the transverse position of the interaction point.
As shown in Fig.~\ref{fig:setup}, 
there are five major magnets between the RPs and the collision point, 
two dipole magnets DX and D0, which bend beams into collision, and 
the focusing triplet of quadrupoles Q1-Q3. The dipole magnets scatter 
out particles with momentum which is not close to the beam momentum.
The detector package inside each RP consists of four 0.4 mm thick silicon micro-strip detector planes  
with a strip pitch of about 100~$\mu$m, two of them measuring the horizontal ($x$) 
and two the vertical ($y$) position of a scattered proton.
The sensitive area of the detectors is 79 $\times$ 48 mm$^2$.
Scintillation counters covering this area  are used to form a trigger for elastic events.
More details on the experiment and the technique can be found in Refs.~\cite{pp2ppplb04, pp2ppNIM}. 

The preliminary alignment was done by 
surveying the detector packages during their assembly and 
after installation inside the Roman Pots with respect to the beam line of the accelerator.
The displacement
of the RPs during data taking was measured by linear variable differential transformers 
(LVDTs). The final alignment was done using elastic events in the 
overlapping regions of horizontal and vertical RPs, which allowed 
a relative position measurement
of the RPs on each side of the IP with a precision better than
0.1~mm. Collinearity of the elastic events and Monte-Carlo  simulations of the acceptance boundaries
due to limiting apertures in the quadrupole magnets were used to further constrain the geometry and to estimate
systematic errors.

The data were taken during four dedicated RHIC stores between June 30 and July 4, 2009 
 with special beam optics of $\beta^*$ = 22~m in order to minimize the angular divergence 
at the IP~\cite{angelika}. 
The average luminosity over the four stores during which the data were collected was 
${\mathcal L}$ $\approx$  2$\cdot$10$^{29}$cm$^{-2}$s$^{-1}$.
The closest approach of the first strip to the center of the beam was about 10~mm or about 12 $\sigma$
of the transverse beam size. 
A total of 33 million elastic triggers were recorded. 

\section{Data selection and reconstruction of elastic scattering events}
The selection of elastic events in this experiment is based on the collinearity of the scattered proton tracks. 
A single track was required on each side of the IP.
Noisy and dead strips were rejected, with a total of five  out of $\approx$ 14,000 in the active detector area.
Track reconstruction started with the search for hits in the silicon detectors. First, adjacent strips with 
collected charge values above $5\:\sigma$ from their pedestal averages were found and combined into clusters. 
A threshold depending on the cluster width was applied to the total charge of the cluster, thus 
improving the signal-to-noise ratio for clusters of 3 to 5 strips, while wider clusters were rejected.
The cluster position was determined as a charge weighted average of strip coordinates. 
For each RP 
a search was performed for matching clusters in the pairs of planes measuring the same coordinate.
Two clusters in such planes were considered matched if the distance between them
was smaller than 200~$\mu$m,  approximately the width of two strips.
A matching pair with the smallest matching distance was chosen and its cluster coordinates were averaged.
If only one cluster in the pair of planes was found, we just use its coordinate for the analysis.
If more than one cluster or no match was found,
no output from this RP was selected.
An $(x,y)$ pair found in an RP was considered a track.
About 1/3 of all reconstructed tracks were found in the region of overlapping acceptance 
between the horizontal and the vertical RPs; for those tracks the average 
of the kinematic variables was used. To minimize the background contribution from 
beam halo particles, products of beam-gas interactions, and
detector noise, fiducial areas were selected to cut edges of the 
silicon detectors near the beam and boundaries of the magnet apertures.

Planar angles $\theta_x^{\rm RP},\theta_y^{\rm RP}$ and coordinates $x^{\rm  RP}, y^{\rm RP}$ of 
protons at a given RP relate to
the angles $\theta_x, \theta_y$ and coordinates $x, y$ at the IP by the transport matrix ${\mathbf M}$:
\begin{linenomath}
\begin{equation}
\begin{bmatrix} 
x^{\rm RP} \\ \theta_x^{\rm RP} \\ y^{\rm RP} \\ \theta_y^{\rm RP}   
\end{bmatrix}
= {\mathbf M} 
\begin{bmatrix} 
x \\ \theta_x\\ y \\ \theta_y   
\end{bmatrix}
=
\begin{bmatrix} 
a_{11} & L_x^{\rm eff} & a_{13} & a_{14}  \\
a_{21} & a_{22} & a_{23} & a_{24}      \\
a_{31} & a_{32} & a_{33} & L_y^{\rm eff}  \\
a_{41} & a_{42} & a_{43} & a_{44}      
\end{bmatrix}
\begin{bmatrix} 
x \\ \theta_x\\ y \\ \theta_y   
\end{bmatrix}
.
\label{eq:tm}
\end{equation}
\end{linenomath}
For example, the transport matrix ${\mathbf M}$ for the horizontal Roman Pot in the West side of the IP (WHI,WHO) is:
\begin{linenomath}
\begin{displaymath}
{\mathbf M} =
\left[
\begin{array}{D{.}{.}{9} D{.}{.}{9} D{.}{.}{9}  D{.}{.}{9} }
-0.0913              &  25.2566~{\rm m}   &  -0.0034               & 0.0765~{\rm m} \\
-0.0396~{\rm m^{-1}} &   0.0137           &  -0.0001~{\rm m^{-1}}  & 0.0057          \\
-0.0033              &   -0.1001~{\rm m}  &   0.1044               & 24.7598~{\rm m} \\
 0.0002~{\rm m^{-1}} &   0.0083           &  -0.0431~{\rm m^{-1}}  & -0.6332 
\end{array}
\right ]
.
\end{displaymath}
\end{linenomath}
For the case of parallel-to-point focusing, and in the absence of $x$-$y$ mixing, the transport matrix 
is simplified and the so-called ``effective" length, $L^{\rm eff}$, terms dominate.
The $L^{\rm eff}$ values are in the range of 22-26~m for this experiment.
The angles of the scattered protons at the IP can then be reconstructed
independently for the East~(E) and West~(W) arms with respect to the IP:
\begin{linenomath}
\begin{eqnarray}
\theta_x &=& x^{\rm RP} / L_x^{\rm eff}, \\
\theta_y &=& y^{\rm RP} / L_y^{\rm eff}.
\label{eq:stm}
\end{eqnarray}
\end{linenomath}
Because non-dominant terms in the transport matrix are small and result in a negligible
correction of about 4 $\mu$rad to the reconstruction of the scattering angles, we used a 2$\times$2 matrix
($L_x^{\rm eff}$, $a_{14}$ ; $a_{32}$, $L_y^{\rm eff}$), which was
obtained by neglecting those small terms of the transport matrix.

\begin{figure}
\includegraphics[width=130mm]{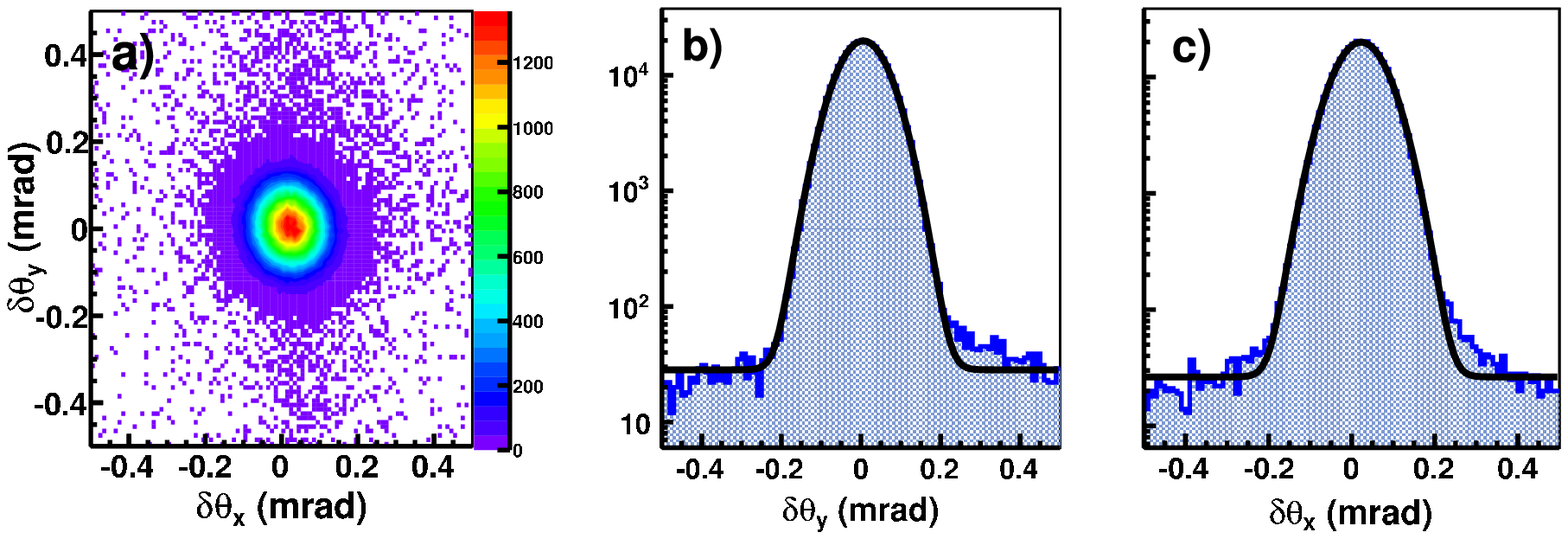}
\caption{(color online) Distribution of $\delta\theta_y$ vs. $\delta\theta_x$ for both
detector pairs in horizontal RPs (a) and their projections in $\delta\theta_y$ (b) and 
$\delta\theta_x$ (c).  
The overlaid curves represent the fits with a Gaussian signal and a linear background.
The $\sigma$ values of distributions are $\approx$ 58 $\mu$rad, consistent with beam angular divergence, and 
the background-to-signal ratio under the Gaussian distributions in ${\pm 3~\sigma}$ is $\approx$ 0.4\%.} 
\label{fig:coll}
\end{figure}

Once the planar angles at IP were reconstructed, a collinearity requirement was imposed using $\chi^2$
defined as:
\begin{linenomath}
\begin{equation}
\chi^2 = {[(\delta\theta_x - {\delta\bar{\theta}_x})/\sigma_{\theta_x}]^2 + 
[(\delta\theta_y-{\delta\bar{\theta}_y})/\sigma_{\theta_y}]^2}\,,  
\label{eq:chi2}
\end{equation}
\end{linenomath}
where $\delta\theta_{x,y} = [\theta_{x,y}^{W} - \theta_{x,y}^{E}]$ 
and the mean values ${\delta\bar{\theta}_{x,y}}$ and widths $\sigma_{\theta_{x,y}}$
are taken from the fits to data performed for each data sample.  An example is shown in Fig.~\ref{fig:coll}.  
The small non-zero mean values ($\approx$ 10 $\mu$rad) are consistent with the uncertainties of
angle determinations discussed in the next section. 
Fig.~\ref{fig:coll} shows a typical distribution of $\delta\theta_y$ vs. $\delta\theta_x$ and its
projections, fitted with a Gaussian and a linear background. 
Based on these fits, the non-collinear background contribution is estimated to be
0.3-0.5\%.
The requirement of $\chi^2<9$ left about 21 million events for the asymmetry calculations.

The polar scattering angle $\theta$ and azimuthal angle $\varphi$ (measured counterclockwise from the positive x-axis) for an event were then calculated
as an average of those obtained from East and West arms, and the four-momentum transfer squared, $t$, was
assigned to the event using $t = -2p^2(1 - {\rm cos}\theta) \approx -p^2\theta^2$
with $p=100.2$~GeV/$c$.

\section{Single Spin Asymmetries}
\label{sec:ssa}
The azimuthal angle dependence of the cross-section for the elastic  collision of 
vertically polarized protons is given~\cite{leader} by:
\begin{linenomath}
\begin{eqnarray}
\frac{d^2\sigma}{dt d\varphi}&=&\frac{1}{2\pi}\frac{d\sigma}{dt}\cdot[1+({\mathcal P}_B+{\mathcal P}_Y)A_N(t){\rm cos}\varphi \nonumber\\
&+&{\mathcal P}_B{\mathcal P}_Y(A_{NN}(t){\rm cos}^2\varphi + A_{SS}(t){\rm sin}^2\varphi)]\;,
\end{eqnarray}
\end{linenomath}
where higher order terms are ignored, $d\sigma/dt$ is the spin-averaged cross-section, ${\mathcal P}_B$ and  ${\mathcal P}_Y$  
are the beam polarizations for the two colliding beams 
(called Blue and Yellow).
The double spin asymmetry $A_{NN}$ is defined as the cross-section asymmetry for
scattering of protons with spin orientations parallel and antiparallel
with respect to the unit vector $\hat{n}$, normal to the scattering plane.
The asymmetry $A_{SS}$ is defined analogously for both
beams fully polarized along the unit vector $\hat{s}$ in the scattering 
plane and normal to the beam.

For each of the four RHIC stores, the event sample satisfying the requirements for elastic 
scattering was divided into five $t$-bins. 
Within each $t$-bin, the $\varphi$ distributions were subdivided into bins of 10$^\circ$. 
The raw asymmetry, $\varepsilon_N(\varphi)$, was calculated using geometric means~\cite{squareroot},
the so-called ``square root formula" for each pair of $\varphi$ and $\pi-\varphi$ bins
in the range $-\pi/2 < \varphi < \pi/2$:
\begin{widetext}
\begin{linenomath}
\begin{equation}
\varepsilon_N(\varphi) = \frac{({\mathcal P}_B+{\mathcal P}_Y)A_N \cos(\varphi)}{1 + \nu(\varphi)} 
=
   \frac{\sqrt{N^{\uparrow\uparrow}(\varphi)N^{\downarrow\downarrow}(\pi-\varphi)} - 
         \sqrt{N^{\downarrow\downarrow}(\varphi)N^{\uparrow\uparrow}(\pi-\varphi)}}%
        {\sqrt{N^{\uparrow\uparrow}(\varphi)N^{\downarrow\downarrow}(\pi-\varphi)} + 
         \sqrt{N^{\downarrow\downarrow}(\varphi)N^{\uparrow\uparrow}(\pi-\varphi)}} \,,
\label{eq:squareroot}
\end{equation}
\end{linenomath}
\end{widetext}
where the ``$\uparrow$" and ``$\downarrow$" indicate the spin direction of the
transversely polarized colliding proton beam bunches, $N$ is the number of events detected in the respective spin and 
respective $\varphi$ states and $\nu(\varphi) = {\mathcal P}_B{\mathcal P}_Y(A_{NN}\cos^2(\varphi)+A_{SS}\sin^2(\varphi))$.

\begin{figure}
\includegraphics[width=92mm]{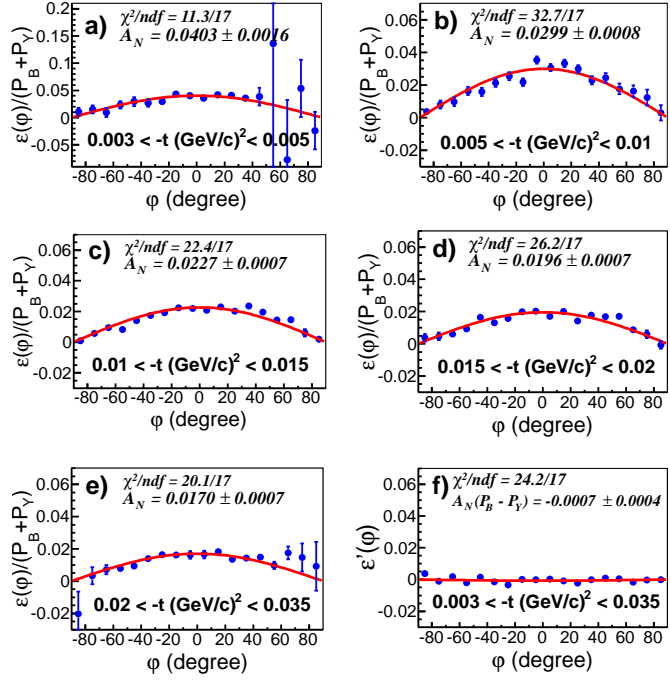}
\caption{(color online) The asymmetry 
$\varepsilon(\varphi)/({\mathcal P}_B+{\mathcal P}_Y)$  
for the five $t$-intervals as given in Table~\ref{tab:results} (a) - (e). 
The asymmetry $\varepsilon'(\varphi)$ for the whole measured $t$-range (f). 
The red curves represent the best fit to Eq.~(\ref{eq:squareroot}) (a) - (e) and Eq.~(\ref{eq:forbidden}) (f).}
\label{fig:rawasymm}
\end{figure}
In the square root formula~(\ref{eq:squareroot}),
the relative luminosities of different spin 
direction combinations cancel out. In addition, the detector acceptance 
and efficiency also cancel out, provided they do not depend on the bunch 
polarization.
Results of Ref.~\cite{doublespin} and preliminary results of this 
experiment~\cite{spin2010} show that both 
$A_{NN}$ and $A_{SS}$ are very small $\approx 0.005$ (and compatible with zero), constraining
$\nu(\varphi)$ to $\approx 0.002$, which can be safely neglected.

For each RHIC store, the obtained raw asymmetries were divided by the sum of polarizations of both beams for 
this particular store, 
and then averaged over the stores. The resulting asymmetries for each $t$ bin
are shown in Fig.~\ref{fig:rawasymm}(a-e) as a function of $\varphi$.
The solid lines represent the best fits to Eq.~(\ref{eq:squareroot}). 

Along with the raw asymmetry, $\varepsilon_N$, which is proportional to the sum of the beam polarizations
$({\mathcal P}_B+{\mathcal P}_Y)$, other asymmetries can be obtained using different combinations 
of bunch spin directions. For instance, the asymmetry proportional to the beam polarization 
difference $({\mathcal P}_B-{\mathcal P}_Y)$ is defined as follows:
\begin{widetext}
\begin{linenomath}
\begin{equation}
\varepsilon'(\varphi) = \frac{({\mathcal P}_B-{\mathcal P}_Y)A_N \cos(\varphi)}{1 - \nu(\varphi)} =
   \frac{\sqrt{N^{\uparrow\downarrow}(\varphi)N^{\downarrow\uparrow}(\pi-\varphi)} - 
         \sqrt{N^{\downarrow\uparrow}(\varphi)N^{\uparrow\downarrow}(\pi-\varphi)}}%
        {\sqrt{N^{\uparrow\downarrow}(\varphi)N^{\downarrow\uparrow}(\pi-\varphi)} + 
         \sqrt{N^{\downarrow\uparrow}(\varphi)N^{\uparrow\downarrow}(\pi-\varphi)}} \; .
\label{eq:forbidden}
\end{equation}
\end{linenomath}
\end{widetext}
Provided that the beam polarizations $({\mathcal P}_B$ and ${\mathcal P}_Y)$ have the
same values, which is approximately valid in  this experiment, one would expect $\varepsilon'=0$. The derived 
values of $\varepsilon'$ may be used to estimate
false asymmetries, which remain after applying the ``square root" method.
The distribution of the asymmetry $\epsilon'$, obtained for the 
whole $t$-range, together with its fit, is shown in Fig.~\ref{fig:rawasymm}(f).

During data taking, 64 bunches (16$\uparrow\uparrow$, 16$\downarrow\downarrow$, 
16$\uparrow\downarrow$, 16$\downarrow\uparrow$) 
of the 90 proton beam bunches collided with usable  spin patterns, and were used for
$\varepsilon_N$
and $\varepsilon'$
calculations.

The major systematic uncertainties of the experiment are due to the
error of the beam polarization measurement, the reconstruction of $t$
and a small background contribution as shown in Fig.~\ref{fig:coll}.
The two main contributions to the uncertainty in the $t$ reconstruction 
are due to the uncertainties of the $L^{\rm eff}$ values and the position of the beam center 
at the RP location. 
The former is mostly due to the uncertainty on values of
the magnetic field strength in the Q1-Q3 focusing quadrupoles,
which is mainly due to uncertainties in the magnet current and field measurements.
The correction to the strength was derived using the correlation between the angle and
position in the RPs for the tracks in the regions where the detector acceptance overlaps. 
An overall correction to the strength of the focusing quadrupoles of 0.5\% was applied. 
The residual systematic error of the field calculation was estimated to be $\approx$ 0.5\%, leading to
$\approx$ 1\% uncertainty in $L^{\rm eff}$ and $\approx$ 1.4\% uncertainty in $t$~\cite{phil}.

The position of the beam center is the reference point for the scattering angle calculations
and effectively absorbs a large set of geometrical unknowns such as beam crossing angles 
and transverse beam positions at the IP, beam shifts from the beam pipe center at the RP location,
as well as survey errors. To accommodate all these uncertainties, corrections to the
survey were introduced based on the comparison of the simulated to the measured $(x,y)$ 
distributions at the horizontal RPs on both sides of the IP.
The simulation of the transport of elastically scattered protons through the RHIC magnets and the apertures
was done and the detector acceptance was calculated. 
The acceptance boundaries from that simulation and the data were compared.
No correction was found for the West side, while for the East side a correction of
($\Delta x, \Delta y$) = (2.5, 1.5) mm was obtained.
The uncertainty of that correction was estimated to be 400 $\mu$m.  After applying that alignment correction,
the collinearity, defined as the average angle difference ${\delta\bar{\theta}_{x,y}}$ 
(see Eq.~(\ref{eq:chi2})), was reduced from $\approx$ 55 $\mu$rad to $\approx$ 10 $\mu$rad.
The remaining alignment uncertainty leads to a value of $\delta t/t$ = 0.0020~[GeV/$c$]$/\sqrt{t}$ and
was added in quadrature to the uncertainty due to $L^{\rm eff}$.
The number of background events in the data is less than 1\% in all $t$-bins 
(e.g. see Fig.~\ref{fig:coll}). Assuming the background is beam polarization independent, the asymmetry
will be diluted by the same amount, $\delta A_N/A_N<0.01$. This  value results
in a negligible contribution to the total error,  when statistical and systematic errors are added in quadrature.

The polarization values of the proton beams were determined by the RHIC
CNI polarimeter group. 
Polarizations and their uncertainties (statistical and systematic combined) for the four stores were: 
0.623$\pm$0.052, 0.548$\pm$0.051, 0.620$\pm$0.053, 0.619$\pm$0.054 (Blue beam),
0.621$\pm$0.071, 0.590$\pm$0.048, 0.644$\pm$0.051, 0.618$\pm$0.048 (Yellow beam)~\cite{cnigroup}. 
The overall luminosity-weighted average polarization values for all four stores are
$\langle {\mathcal P}_B+{\mathcal P}_Y \rangle$~=~1.224$\pm$0.038 and 
\mbox{$\langle {\mathcal P}_B-{\mathcal P}_Y \rangle$~=~$-$0.016$\pm$0.038. }
Taking into account the overall uncertainty for normalization in polarization measurements, 
the total polarization error
$\delta\langle{{\mathcal P}_B+{\mathcal P}_Y}\rangle/$$\langle {\mathcal P}_B+{\mathcal P}_Y \rangle$ is 5.4\%.

If the false asymmetry $\varepsilon_F$ were proportional to the beam polarization values, 
it would be indistinguishable from $A_N$.
On the contrary, if it does not depend on the polarization, it contributes equally to both 
$\varepsilon_N$ and $\varepsilon'$:
\begin{linenomath}
\begin{eqnarray}
\varepsilon_N &=& A_N({\mathcal P}_B+{\mathcal P}_Y)+\varepsilon_F \;,\\
\varepsilon' &=& A_N({\mathcal P}_B-{\mathcal P}_Y)+\varepsilon_F \; .\,
\end{eqnarray}
\end{linenomath}
and a direct estimate on the false asymmetry can be obtained:
\begin{linenomath}
\begin{equation}
\varepsilon_F=\frac{\varepsilon'({\mathcal P}_B+{\mathcal P}_Y)-\varepsilon_N({\mathcal P}_B-{\mathcal P}_Y)}
                  {2{\mathcal P}_Y}
\approx\varepsilon'-\varepsilon_N\frac{{\mathcal P}_B-{\mathcal P}_Y}{{\mathcal P}_B+{\mathcal P}_Y} \,.
\end{equation}
\end{linenomath}
The values of the raw asymmetries, measured in the whole $t$-range, are
$\varepsilon_N$ = 0.0276$\pm$0.0004 and
$\varepsilon'$ = $-$0.0007$\pm$0.0004. This gives a false asymmetry of $\varepsilon_F$  = $-$0.0004$\pm$0.0010.
Thus the conclusion is that the false asymmetry is consistent with zero and very small compared
to the measured raw asymmetry $\varepsilon_N$.

The results of the $A_N$ measurements in the five $t$-bins are summarized in Table~\ref{tab:results} 
together with associated uncertainties and $-t$ range boundaries. Two independent analyses 
of the data performed  with slightly different selection criteria
by two different groups gave consistent results.  
We have also done the cross checks to extract $A_N$ using the beam polarizations of the two beams.  The resulting $A_N$ were found to be compatible with those in Table~\ref{tab:results} within their statistical uncertainties.
\begin{table}[htbp]
   \centering
   \begin{tabular}{@{} lccccc @{}} 
      \toprule
     \hline
      \cmidrule(r){1-2} 
      $-t$  [(GeV/$c$)$^2$]   & ~0.003 - 0.005& ~0.005 - 0.01   & ~0.01 - 0.015 &  ~0.015 - 0.02  & ~0.02 - 0.035 \\
      \midrule
      \hline
      No. of Events  &  444045 & 2091977 & 2854764 & 2882893 & 2502703 \\ 
      \hline
      $\langle-t\rangle$  [(GeV/$c$)$^2$]   & 0.0039        & 0.0077       & 0.0126        &  0.0175 & 0.0232 \\
      $\delta t$  [(GeV/$c$)$^2$](syst.)    & 0.0001        & 0.0002        & 0.0003         &  0.0004          & 0.0004  \\
      \hline
      $A_N$                               & 0.0403        & 0.0299         & 0.0227        &  0.0196            & 0.0170 \\
      $\delta A_N$(stat.)                 & 0.0016        & 0.0008         & 0.0007        &  0.0007            & 0.0007 \\
      $\delta A_N$(syst.)                  & 0.0021        & 0.0016         & 0.0012        &  0.0010            & 0.0009 \\
      \hline
       \bottomrule
   \end{tabular}
   \caption{$A_N$ values in five $t$ ranges with associated uncertainties. Statistical errors for $t$ are negligible and
   combined systematic errors are shown (See the text for details). Statistical errors and systematic errors on $A_N$ are also shown, where
$\delta A_N$(syst.) is a scale error due to the beam polarization. }
   \label{tab:results}
\end{table}

\begin{figure}
\includegraphics[width=88mm]{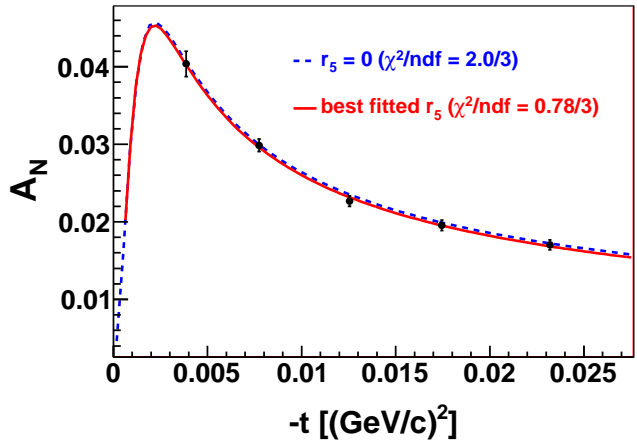}
\caption{(color online) The measured single spin asymmetry $A_N$ for five $-t$ intervals. Vertical
error bars show statistical uncertainties. Statistical error bars in $-t$ are smaller than the plot symbols. 
The dashed curve corresponds to theoretical calculations 
without hadronic spin-flip and the solid one represents the $r_5$ fit.  
}
\label{fig:an}
\end{figure}
\begin{figure}
\includegraphics[width=80mm]{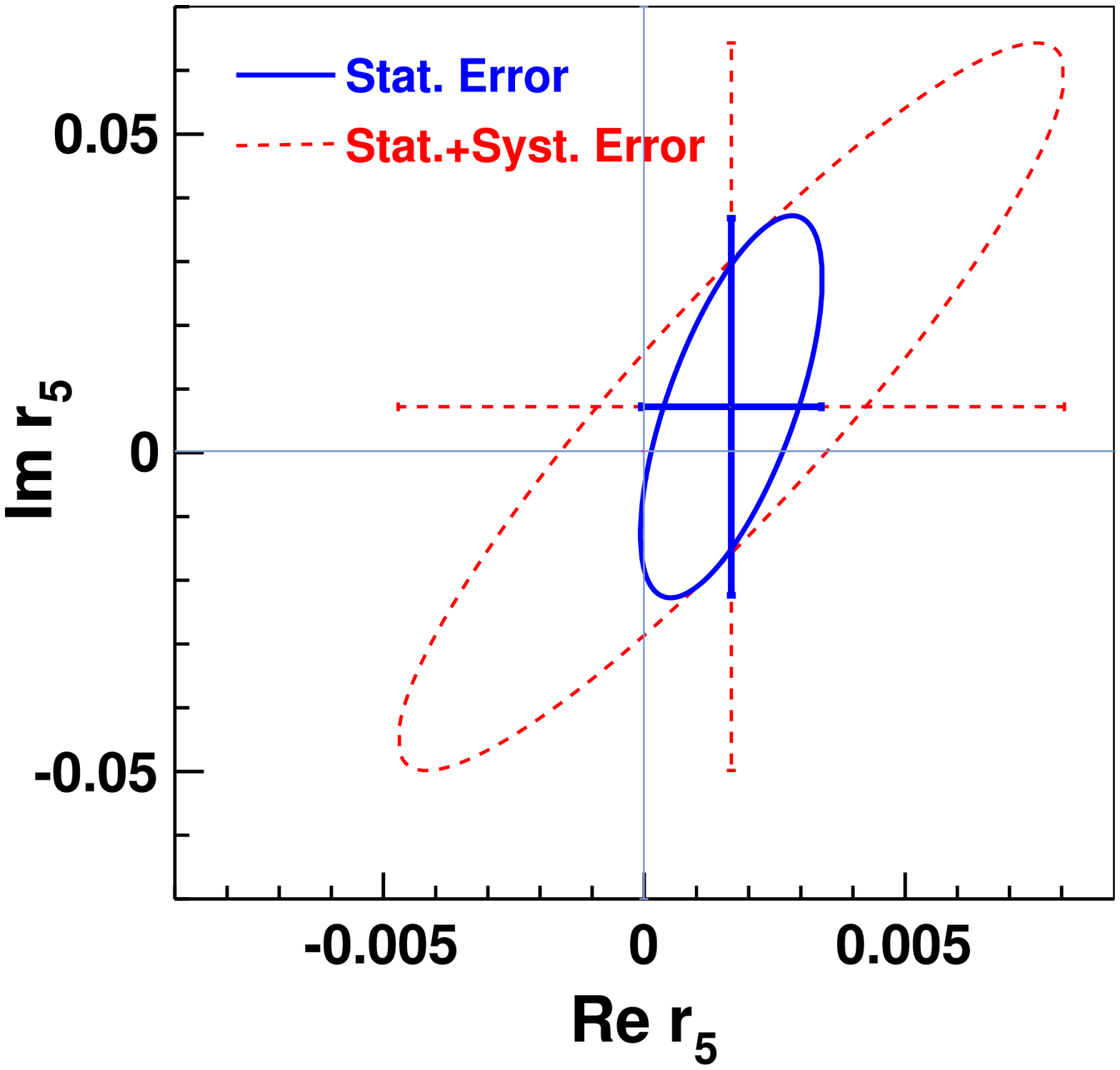}
\caption{(color online) Fitted value of $r_5$ with contours corresponding to statistical error only (solid ellipse and cross) and statistical+systematic errors (dashed ellipse and cross) of $1\sigma$. }
\label{fig:r5}
\end{figure}

\section{Results and Conclusions}

The measured values of $A_N$ 
are shown in Table~\ref{tab:results} and presented in Fig.~\ref{fig:an}  
together with parameterizations based on formula~(\ref{cnicurve}): the dashed line corresponds to no
hadronic spin-flip contribution, i.e. $r_5=0$, while the solid line is the result
of the fit using $r_5$ as a free parameter. 
Other parameter values used in the fit are: $\sigma_{\rm total} = 51.79\pm$0.12~mb, $\rho = 0.1278\pm0.0015$ 
taken from fits to the world pp and ${\rm p}\overline{{\rm p}}$ data~\cite{compete,bourrely2} and 
$B=16.3\pm$1.8~(GeV/$c$)$^{-2}$ from Ref.~\cite{pp2ppplb04}.

The value of $r_5$ resulting from the fit described above is shown in Fig.~\ref{fig:r5}
together with 1$\sigma$ confidence level contours.
 \begin{table}[htbp]
    \centering
    \begin{tabular}{@{} lcrr @{}} 
       \toprule
      \hline
       \cmidrule(r){1-2} 
              & central value    & Re$\:r_5$=0.0017 &  Im$\:r_5$=0.007 \\
              \hline
               \hline
               & uncertainties    & $~~~\delta$Re$\:r_5$ &  ~~~$\delta$Im$\:r_5$ \\
       \midrule
       \hline
       1 & statistical      & 0.0017 &   0.030  \\
       2 &  $\delta t$($L^{\rm eff}$)    & 0.0008 & 0.005    \\
       3 &  $\delta t$(alignment)         & 0.0011 & 0.011    \\
       4 &  ${\delta\mathcal P}$          & 0.0059 & 0.047    \\
              \hline
       5 &  $\delta\sigma_{\rm total}$       & 0.0003 & 0.002    \\ 
       6 &  $\delta\rho$                        & $<$ 0.0001 & $<$ 0.001    \\  
       7 &  $\delta B$                            & $<$ 0.0001 & $<$ 0.001    \\   
             \hline
          & total syst. error                     & 0.0061 & 0.049    \\
          & total stat. + syst. error          & 0.0063 & 0.057    \\
       \hline
        \bottomrule
    \end{tabular}
    \caption{The fitted $r_5$ values including the uncertainties. (1): Statistical uncertainties. (2)-(4): Systematic uncertainties associated with this measurement. 
(5)-(7): Systematic uncertainties associated with 
    the values used in the fit function. See the text for details. }
    \label{tab:systematics}
 \end{table}
\begin{figure}
\includegraphics[width=80mm]{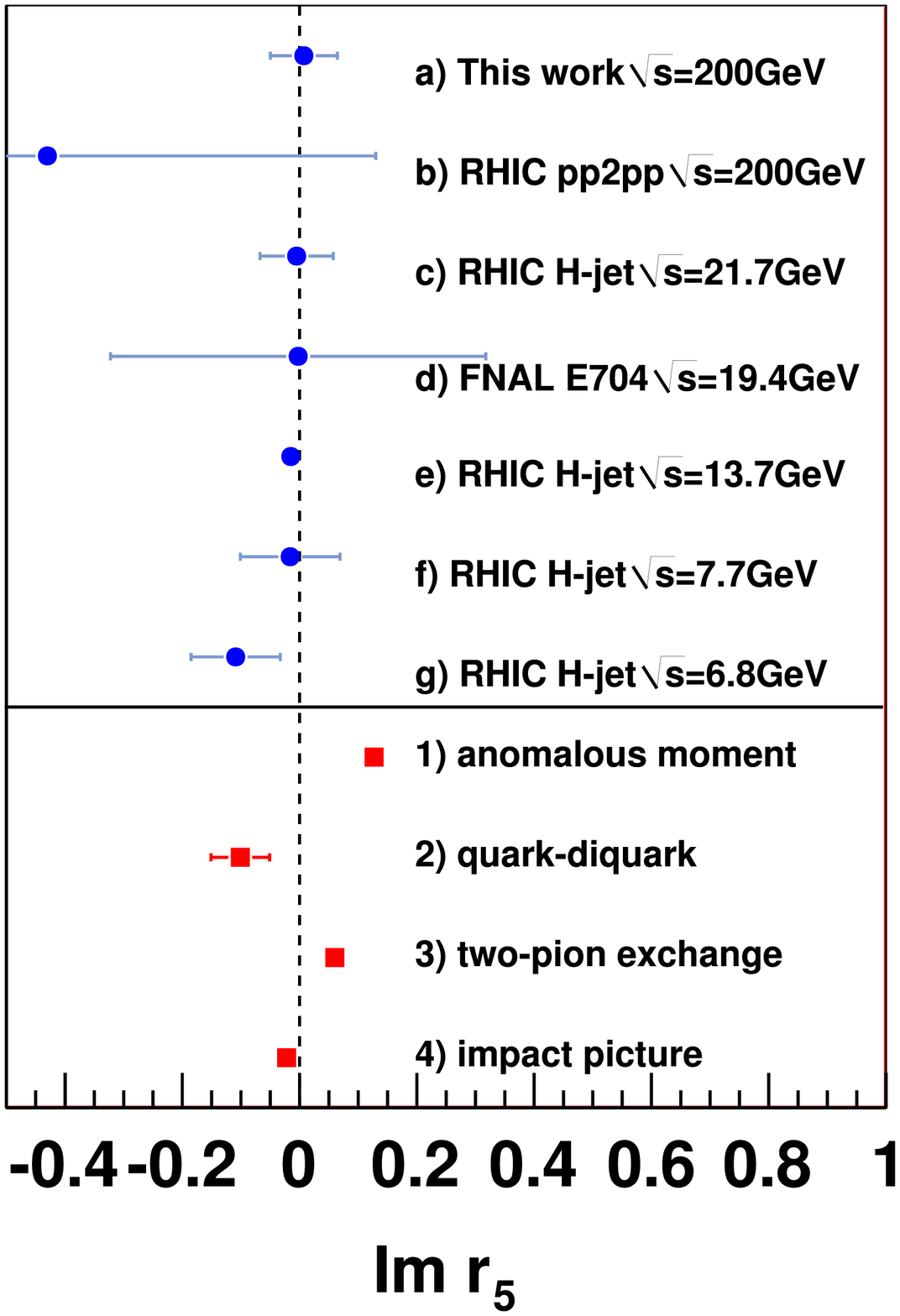}
\caption{(color online) Measurements of  Im($\:r_5$) values for (a) this experiment, 
(b) RHIC pp2pp at $\sqrt{s}$=200 GeV~\cite{pp2ppAN}, 
(c) RHIC H-jet target at $\sqrt{s}$=21.7 GeV~\cite{bazilevsky},  
(d) FNAL E704 at $\sqrt{s}$=19.4 GeV~\cite{e704} , 
(e) RHIC H-jet target at $\sqrt{s}$=13.7 GeV~\cite{jetcal}, 
(f) RHIC H-jet target at $\sqrt{s}$=7.7 GeV~\cite{bazilevsky}, and  
(g) RHIC H-jet target at $\sqrt{s}$=6.8 GeV~\cite{alekseev}.  
Theoretical calculations shown are 
(1) anomalous moment~\cite{ryskin}, (2) quark-diquark picture~\cite{diquark}, 
(3) two-pion exchange model~\cite{pumplin}, and
(4) impact picture~\cite{bourrely}. The theoretical calculations are either energy independent (1,2,3) or done
at $\sqrt{s}$=200 GeV (4). The vertical dashed line indicates where Im($\:r_5$)=0. 
All error bars shown include both statistical and systematic errors.
}
\label{fig:r5compare}
\end{figure}
In Table~\ref{tab:systematics}, we show the central value of the fit  and uncertainties on Re$\:r_5$ and Im$\:r_5$ 
due to the listed effects. 
In the first row of the table, the statistical error to the fit with the central value
of the parameters is shown. The remaining rows show changes of Re$\:r_5$ and Im$\:r_5$,
when each parameter was varied one by one by $\pm$1$\:\sigma$ during the fit procedure. 
Rows 2 and 3 show the effect due to the systematic uncertainty in $L^{\rm eff}$ and alignment, 
row 4 due to the beam polarization (vertical scale uncertainty of $A_N$) and rows 5-7 systematic 
contributions due to the uncertainty of fit parameters. The dominant source of  
the systematic uncertainty is due to the beam polarization uncertainty. 
The total systematic uncertainty, including the effects related to rows 2-7 of Table~\ref{tab:systematics}, is obtained by 
adding the error covariance matrices. 

The final result on $r_5$ is shown in Fig.~\ref{fig:r5}
together with both statistical and systematic uncertainties.
The obtained values Re$\:r_5$ = 0.0017$\pm$0.0063 and Im$\:r_5$ = 0.007$\pm$0.057 are
consistent with the hypothesis of no hadronic spin-flip contribution
at the energy of this experiment. 

Since the maximum $A_N$ in the CNI region can be evaluated as $\kappa - 2{\rm Im}\:r_5$ 
in Eq.~(\ref{cnicurve}), theoretical calculations emphasize values of Im$\:r_5$.  
Measurements of Im$\:r_5$ at different energies in the range 6.8 GeV $\leqslant \sqrt{s} \leqslant$ 200 GeV 
are shown in Fig.~\ref{fig:r5compare}, together with predictions of theoretical models 
of the hadronic spin-flip amplitude as discussed above. All of the experimental
results, including that reported here,  are consistent with the assumption of no
hadronic spin-flip contribution to the elastic proton-proton scattering. 
The   high accuracy of the current measurement provides strong limits on 
the size of any hadronic spin-flip amplitude at this high energy, hence
significantly constraining theoretical models which require hadronic spin-flip.

\bibliography{apssamp} 

\begin{acknowledgments}
We thank the RHIC Operations Group and RCF at BNL, the NERSC Center at LBNL and the Open Science Grid consortium for providing resources and support. This work was supported in part by the Offices of NP and HEP within the U.S. DOE Office of Science, the U.S. NSF, the Sloan Foundation, CNRS/IN2P3, FAPESP CNPq of Brazil, Ministry of Ed. and Sci. of the Russian Federation, NNSFC, CAS, MoST, and MoE of China, GA and MSMT of the Czech Republic, FOM and NWO of the Netherlands, DAE, DST, and CSIR of India, Polish Ministry of Sci. and Higher Ed., National Research Foundation (NRF-2012004024), Ministry of Sci., Ed. and Sports of the Rep. of Croatia, and RosAtom of Russia.
\end{acknowledgments}

\end{document}